\documentstyle[twoside,fleqn,espcrc2,epsfig]{article}

\newcommand{\aive}{a\"\i{}ve }

\newcommand{\half}{{\textstyle{\frac{1}{2}}}}

\newcommand{\imag}{{\rm i\hspace{0.13ex}}}

\newcommand{\bgeq}{\begin{equation}}
\newcommand{\bgeqa}{\begin{eqnarray}}
\newcommand{\edeq}{\end{equation}}
\newcommand{\edeqa}{\end{eqnarray}}

\newcommand{\PRD}[1]{Phys.\ Rev.\ \textbf{D{}#1}}

\title{Mass renormalisation for improved staggered quarks}

\author{J.\ Hein\address{Department of Physics and Astronomy, University of Edinburgh, EH9 3JZ, Scotland, UK},
Q.\ Mason\address{%
Newman Laboratory of Nuclear Studies, Cornell University, Ithaca, NY 14850, USA},
G.P.~Lepage$^{\rm b}$, 
H.~Trottier\address{Physics Department, Simon Fraser University, Burnaby,
B.C., Canada V5A 1S6}}

\begin{document}

\begin{abstract}
  Improved staggered quark actions are designed to suppress flavour
  changing strong interactions. We discuss the perturbation theory for
  this type of actions and show the improvements to reduce the
  quark mass renormalisation compared to n\aive
  staggered quarks. The renormalisations are of similar size as 
  for Wilson quarks.
\end{abstract}

\maketitle 

\noindent\hspace*{-1.9mm}
\raisebox{8cm}[0ex][0ex]{
{\normalsize
\parbox{4cm}{
\textbf{\textsf{Edinburgh 2001/17}}\\
\textbf{\textsf{CLNS 01/1757}}\\
\textbf{\textsf{hep-lat/0110045}}}}
}\vspace*{-5ex}

\section{INTRODUCTION}
The n\aive discretisation of the fermionic action leads to additional
poles in the corners of the Brillioun zone. Such a theory describes
several flavours of degenerate quarks propagating through the
lattice. In the case of gauge interactions this leads to flavour changes
by exchange of highly virtual gauge bosons with one or several
momentum components $aq_\mu \approx \pi$. For QED and QCD
flavour changing interactions are unphysical.  
These interactions lift the degeneracy of the pions
associated with the different quark flavours. They also lead to large
renormalisations, originating from the high
$q$-region in loop integrals involving fermion propagators. 
Due to the doubling, for high $q$ the fermions are almost on-shell, 
leading to excessive  contributions from this momentum region.

The staggered quark action reduces the doubling without completely
eliminating it.  Here we like to discuss the effect of improved staggered
quark actions with suppressed flavour changing interactions on the
renormalisation of the mass of staggered quarks. For a discussion on
the splitting of the pions see for example D.~Toussaint's review
\cite{toussaint} and the references therein.

\section{FAT LINKS} 
By fattening the links in the action, the coupling of the 
quarks to highly virtual gluons can be suppressed \cite{blum,flsym}.
This is part of the Symanzik improvement program for staggered quarks.
The gauge link variables $U_{x,\nu}$ have to be replaced with a fat-link
$V_{x,\nu}$; $\Delta_\nu$ denoting a covariant second lattice derivative
\bgeq\label{veq}
V_{x,\mu} := 
\prod_{\nu\ne\mu}\left.\left(1 + \frac{\Delta_\nu}{4}\right)
         \right|_{\rm symm.} U_{x,\mu}\,.
\edeq
This replacement removes the coupling to high momentum gluons at
tree-level. 
%
%
%
The fat links introduce a flavour conserving ${\mathcal{O}}(a^2)$
artifact. This can be removed by further replacing $V_{x,\mu} \to
V_{x,\mu}'$ 
\bgeq \label{lecorr}
V_{x,\mu}'  = V_{x,\mu} - \frac14 \sum_{\nu\atop\nu\ne\mu}
(\nabla_\rho)^2 U_{x,\mu}\,.
\edeq
The ${\mathcal{O}}(a^2)$ improvement of the staggered quark action is
completed by the addition of the Naik term to the derivative. Our
results including the Naik term are still preliminary.
Further improvements removing flavour changing
interactions from the action at ${\mathcal{O}}(\alpha_s)$ are reported in
\cite{paulposter}.

\section{QUARK-ANTIQUARK-GLUON VERTEX}
\begin{figure}
\centerline{\epsfig{file=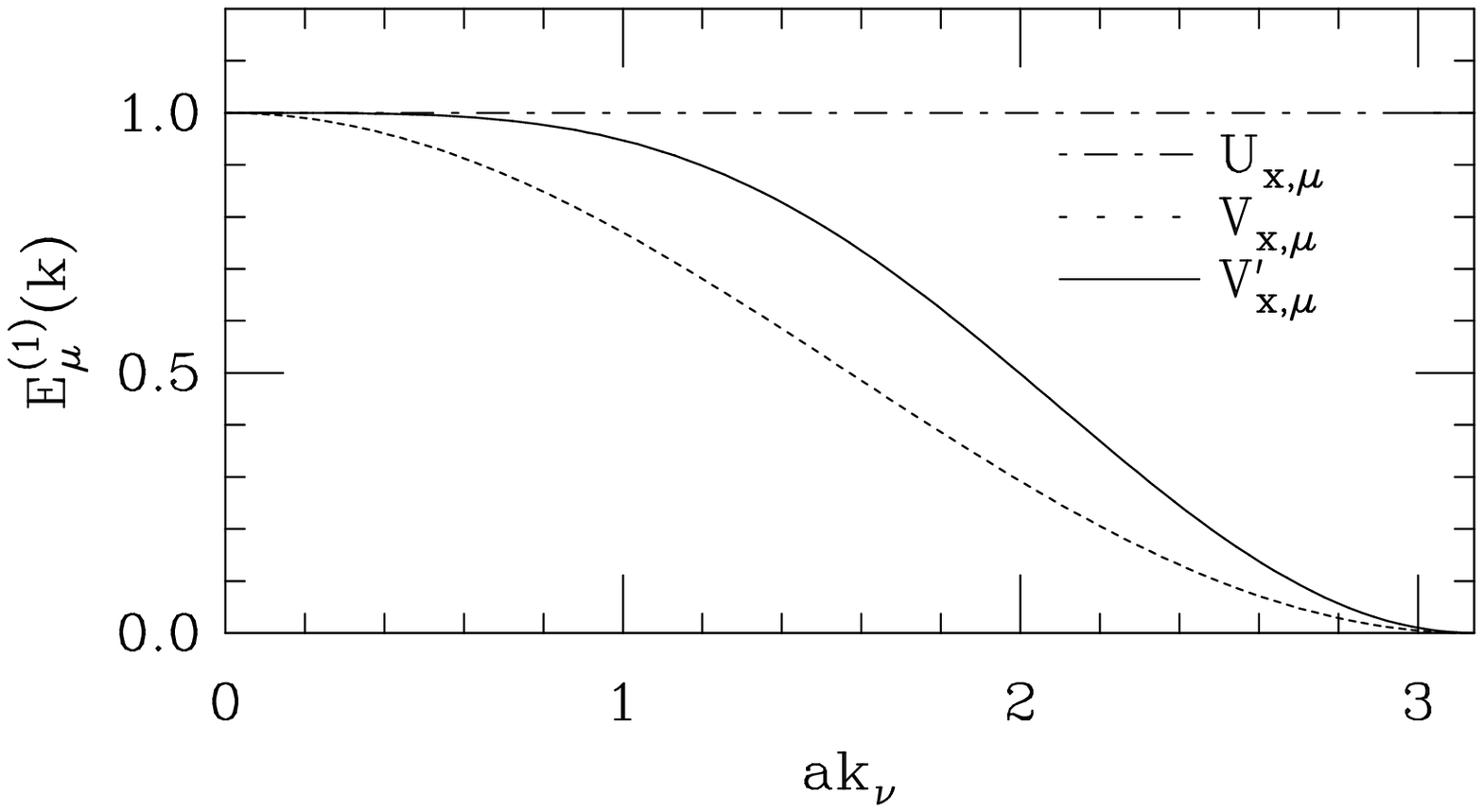,width=6cm,bburx=507pt,bbury=303pt,
bbllx=38pt,bblly=37pt}}
\centerline{\epsfig{file=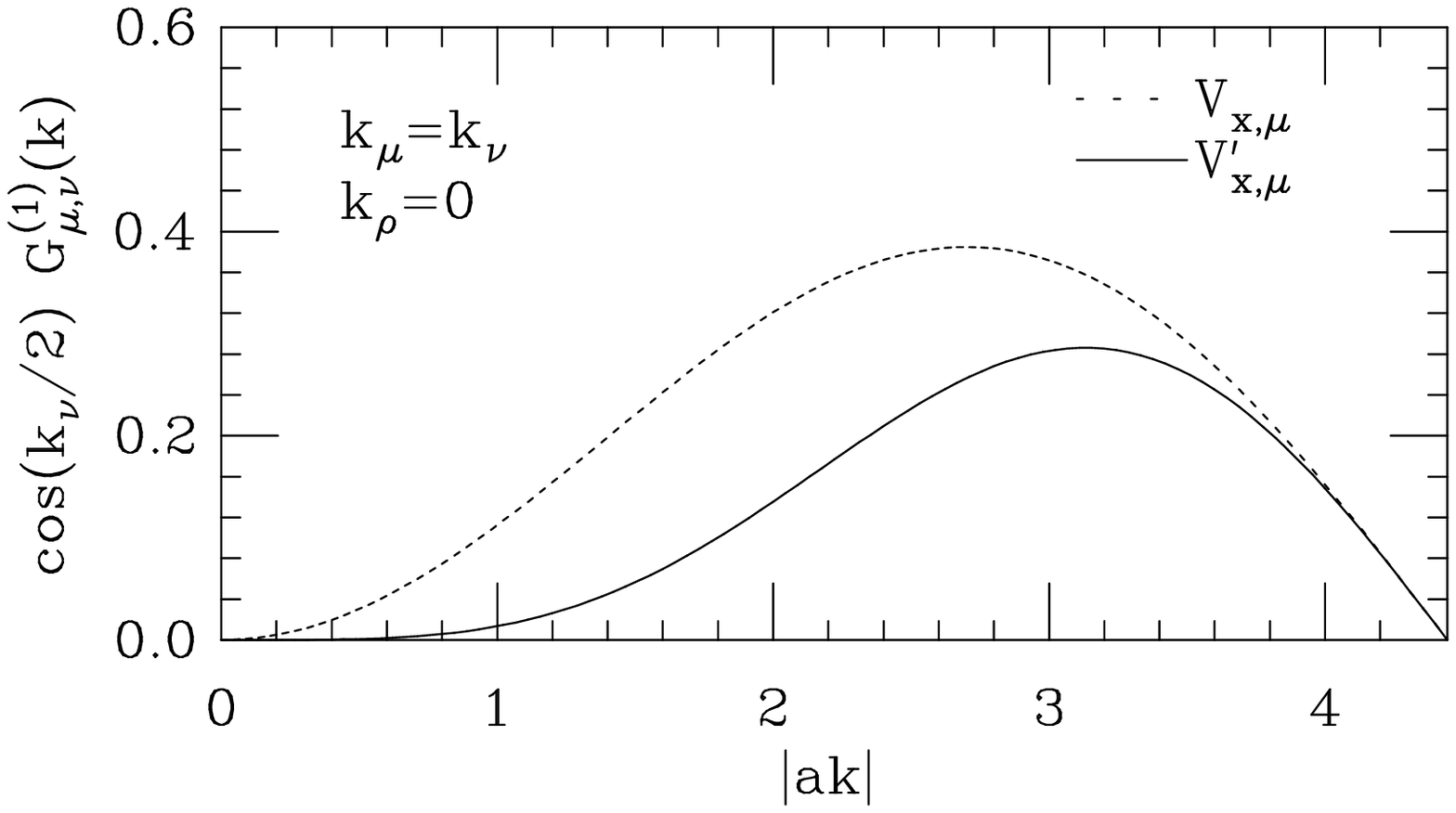,width=6cm,bburx=507pt,bbury=303pt,
bbllx=38pt,bblly=47pt}}
\vspace*{-10mm}
\caption{\label{coeffig}Coeffcient functions for the single gluon vertex}
\end{figure}

The Feynman rules display nicely how the fattening suppresses the
coupling between the quark-antiquark pair and a single gluon with
large transverse momentum. Let us consider the quark-antiquark-gluon vertex
\bgeqa 
{\cal V}_1 &=&
\imag g \Bigg[
\gamma_\mu \cos (\half(p+q)_\mu) E^{(1)}_\mu(q-p)\,, \nonumber \\ 
&&
 + \sum_{\nu \ne \mu} \gamma_\nu \cos (\half(p+q)_\nu)\,
G^{(1)}_{\mu,\nu}(q-p)
\Bigg]T^b\,. \nonumber \\
\edeqa
For the n\aive action one gets $E^{(1)}_\mu=1$ and
$G^{(1)}_{\mu,\nu}=0$ independent of momentum. For fat links according
to eq.~(\ref{veq}) we obtain
\bgeqa
E^{(1)}_\mu(k)&=& \prod_{\nu \atop \nu\ne\mu} \cos^2(\half k_\nu)\,,\\
G^{(1)}_{\nu,\mu}(k) &=& \sin(\half k_\nu)\sin(\half k_\mu)
\Bigg[
\frac13\!\prod_{\sigma \atop \sigma\ne\mu,\nu}\!\!\cos^2(\half k_\sigma)
\nonumber \\
&&+\frac{1}{6}\!\sum_{\sigma \atop\sigma\ne\mu,\nu}\!\cos^2(\half k_\sigma)
+ \frac13
\Bigg]\,.
\edeqa
Including the improvement term in eq.~(\ref{lecorr}) this becomes
\bgeqa
E^{(1)}_\mu{}'(k) &=& E^{(1)}_\mu{}(k) + \frac14
 \sum_{\nu\atop\nu\ne\mu}\sin^2(k_\nu)\,,\\
G^{(1)}_{\sigma,\mu}{}'(k) &=& G^{(1)}_{\sigma,\mu}{}(k) \nonumber\\
&&
- \frac12 \sin(\half k_\mu) \sin(k_\sigma) \cos(\half k_\sigma)\,.
\edeqa
Fig.~\ref{coeffig} shows the momentum behaviour of the coefficient
functions $E^{(1)}_\mu$ and $G^{(1)}_{\sigma,\mu}$ for the different
actions. In the case of fat links $E^{(1)}_\mu \approx 1$ for small
transverse momenta, which  is the physically relevant region. For the
large flavour changing momenta $ak_\nu\approx \pi$ the coupling
vanishes as intended. The figure also shows the improvements for
$k_\nu \approx 0$ arising from using $V'_{x,\mu}$.

The coefficient $G^{(1)}_{\sigma,\mu}$ describes an unphysical
coupling between a gluon field $A_\sigma$ to the $\gamma_\mu$-term in
the action. Such terms arises from the sides of the staples when
fattening the links. Fortunately this coupling vanishes in the
physical region around $k=0$ as well as in the corner of the
Brillioun zone. Flavour changing interactions are not
reentering through the back door. Again $V'_{x,\mu}$ gives a
substantial improvement for $k\approx 0$.
We also worked out the Feynman rules for the
quark-antiquark-gluon-gluon Vertex as needed for the tadpole diagram.

\section{MASS RENORMALISATION}
\begin{figure}
\centerline{
\epsfig{file=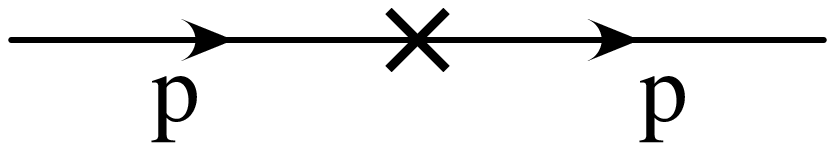,width=0.7in,
bbllx=189pt,bburx=423pt,bblly=300pt,bbury=408pt}\hspace{0.5ex}
\epsfig{file=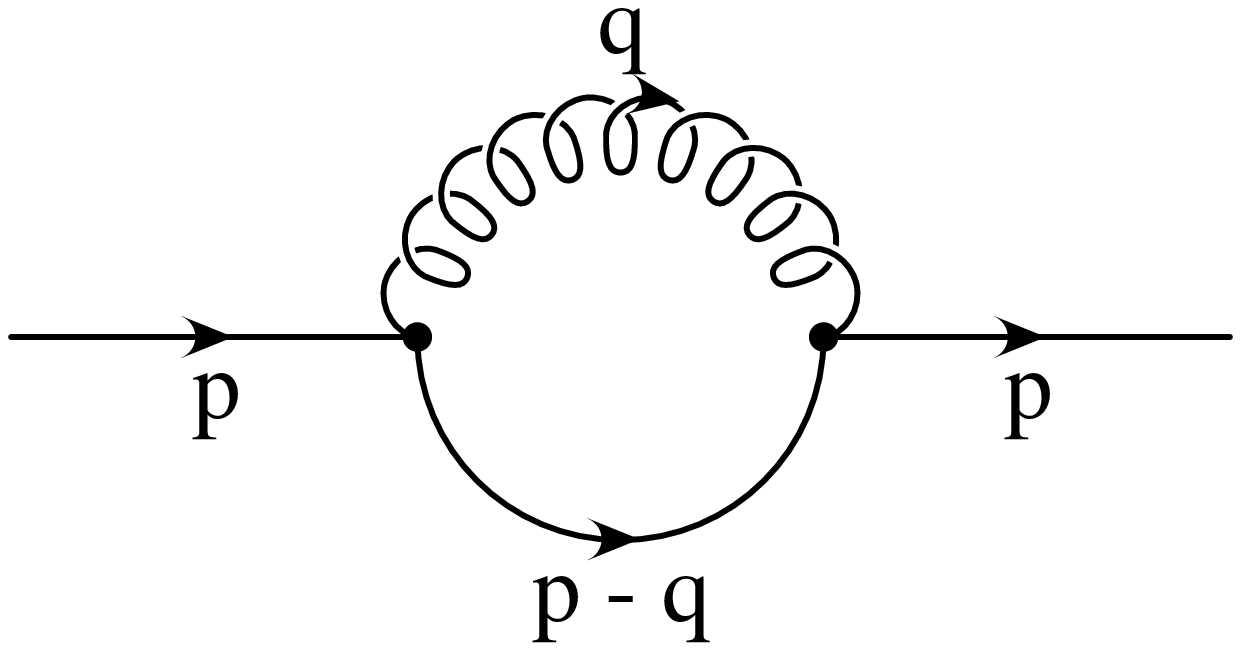,width=1.05in}\hspace{0.5ex}
\epsfig{file=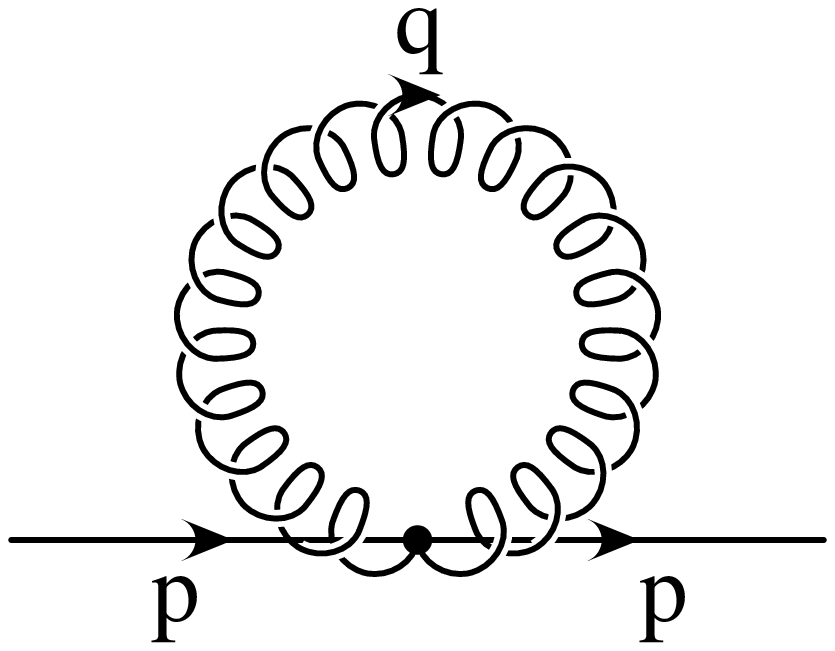,width=0.7in}
}
\vspace*{-10mm}
\caption{\label{feyndiagsfig}Feynman diagrams for the self-energy:
  counter term, rainbow and tadpole diagram}
\end{figure}

We use the following definitions to relate the bare mass $m$ in the
action and the renormalised pole mass $m_{\rm pole}$
\bgeq
m_{\rm pole}
= m\left(1 +  \frac{\delta m^{(0)}}{m} +
\frac{g^2}{4\pi} \frac{\delta m^{(2)}}{m} + 
{\mathcal O}(g^4)\right).
\edeq
For the self energy
$\Sigma^{(2)}$ we use the on-shell condition $p_0=\imag (m + \delta
m^{(0)})$, $p_m=0$. 
The diagrams for $\Sigma^{(2)}$ are shown in
fig.~\ref{feyndiagsfig}. 
The rainbow diagram is the one which is affected by the flavour
changing interactions.

Tadpole improvement \cite{tadimp} turns out to be
crucial for the fat-link actions. In Feynman gauge we observe an
increased tadpole contribution for fat links. This is always
matched by similar sized tadpole improvement counter term of
opposite sign. To achieve this cancellation, tadpole improvement has to be
implemented after working out the higher derivatives of the link operators in the action \cite{mpaper}.
The total contribution of these two graphs to $\delta
m^{(2)}/m$ was about $\half$ or smaller. 

\begin{figure}
\centerline{\epsfig{file=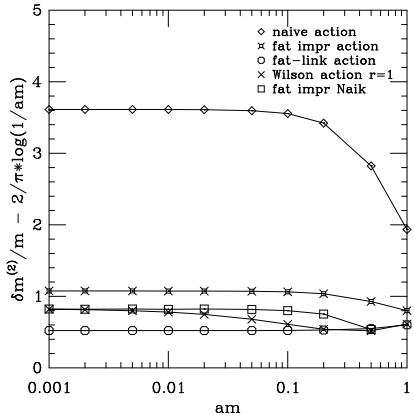,width=6cm}}
\vspace*{-10mm}
\caption{\label{mshiftfig} Mass shift $\delta m^{(2)}$ for 
different actions. }
\end{figure}

For $am \to 0$ the mass shift $\delta m^{(2)}/m$ diverges as
$\frac{2}{\pi}\log(\frac{1}{am})$. Our result for $\delta m^{(2)}/m$
is presented in fig.~\ref{mshiftfig}, where the above divergence is
subtracted.
We used tadpole improvement as defined from the average link in Landau
gauge and the gluon propagator from the Wilson gauge action.
The n\aive action leads to the well known large renormalisation,
which
is substantially  reduced by the introduction of the fat links. The figure
also shows the effect of including eq.~(\ref{lecorr}) and the Naik
improvement term.
These terms have little effect on the result compared to the
fattening of the links. The figure also includes standard Wilson quarks. Their
renormalisations are shown to be of similar size as for the fat-link
quarks. This comparison clearly shows flavour changing interactions to
be the cause of the large renormalisations.

In this context it is interesting to have a look at the momentum scale
$aq^*$ \cite{tadimp}, which can be understood as the ``average loop
momentum''. One observes a significantly larger $q^*$ for the n\aive
action compared to all other actions. This larger $q^*$ can be traced
to the flavour changing interactions in the rainbow diagram.
\begin{figure}
\centerline{\epsfig{file=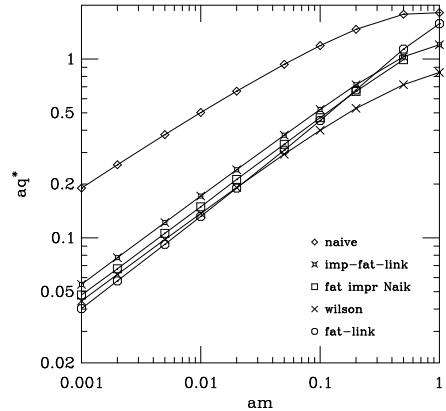,width=6cm}}
\vspace*{-10mm}
\caption{\label{qstarfig} Momentum scale $aq^*$ for different actions. 
}
\end{figure}

\section{CONCLUSION AND OUTLOOK}
Improved staggered quarks suppress strong flavour changing
interactions by fattening the links. This improvement reduces the
known large mass renormalisations for n\aive staggered quarks to the
same level as for Wilson quarks. This will allow for a reliable
calculation of quark masses using staggered quarks in the future.
Preliminary results for current and four quark operators indicate
reduced renormalisations, when using improved staggered quarks
\cite{paulposter}.

\subsection*{Acknowledgements}
J.H. is supported by the European Community's Human potential
programme under HPRN-CT-2000- 00145 Hadrons/LatticeQCD. This work is
supported in part by a grant from the National Science Foundation.

\end{document}